\documentclass[prx,floatfix,twocolumn,showpacs]{revtex4-1}
\usepackage{amsmath}
\usepackage{graphicx}
\graphicspath{{/home/rajeev/Dropbox/Shared/dibyendu/free_fermions_long_range_hopping/}}
\usepackage[colorlinks,linkcolor=blue,citecolor=blue,urlcolor=blue]{hyperref}
\usepackage[usenames,dvipsnames]{color}

\newcommand{\beq}{\begin{equation}}
\newcommand{\eeq}{\end{equation}}
\newcommand{\bes}{\begin{subequations}}
\newcommand{\ees}{\end{subequations}}
\newcommand{\bea}{\begin{eqnarray}}
\newcommand{\eea}{\end{eqnarray}}
\newcommand{\ba}{\begin{array}}
\newcommand{\ea}{\end{array}}
\newcommand{\beqn}{\begin{eqnarray*}}
\newcommand{\eeqn}{\end{eqnarray*}}

\newcommand{\f}[2]{\frac{#1}{#2}}

\newcommand{\ra}{\rangle}

\newcommand{\dg}{\dagger}

\begin{document}
\title{Effect of long-range hopping and interactions on entanglement dynamics and many-body localization}
\author{Rajeev Singh$^1$, Roderich Moessner$^2$ and Dibyendu Roy$^3$}
\address{$^1$Department of Physics, Jack and Pearl Resnick Institute, Bar-Ilan University, Ramat-Gan 52900, Israel}
\address{$^2$Max Planck Institute for the Physics of Complex Systems, N{\"o}thnitzer Str. 38, 01187 Dresden, Germany}
\address{$^3$Raman Research Institute, Bangalore 560080, India}
\date{\today}
\begin{abstract}
We numerically investigate the dynamics of entanglement in a chain of spinless fermions with nonrandom but long-range hopping and interactions, and with random on-site energies. For moderate disorder in the absence of interactions, the chain hosts delocalized states at the top of the band  which undergo a delocalization-localization transition with increasing disorder. We find an interesting regime in this noninteracting disordered chain where the long-time entanglement entropy scales as $S(t) \sim \ln t$ and the saturated entanglement entropy scales with system size $L$ as $S(L,t \to {\infty}) \sim \ln L$.
We further study the interplay of long-range hopping and interactions on the growth of entanglement and the many-body localization (MBL) transition in this system.
We develop an analogy to higher-dimensional short-range systems to compare and contrast such behavior with the physics of MBL in a higher dimension.
\end{abstract}
\maketitle
\section{Introduction} \label{intro}
Recent studies of the time-evolution (dynamics) of entanglement after a global quench from a product state in various disordered quantum systems have enriched our understanding of the physics of localization~\cite{DeChiara06, Znidaric2008, Bardarson2012, IgloiPRB12, Vosk2013, Pino14, Singh2015, Zhao16, DeTomasi2016}.
For example, dynamics of entanglement shows different features in the many-body localized (MBL) phase compared to the Anderson localized (AL) phase, although dc-transport measurements are insensitive to the difference between the two~\cite{Basko2006, Bardarson2012}.
 Dynamical entanglement also changes its nature across the transition between a delocalized and an MBL phase of a disordered interacting quantum system, thereby providing a tool to separate the MBL phase from the noninteracting AL and interacting delocalized phases~\cite{Znidaric2008, Bardarson2012, Singh2015}.

Dynamics of entanglement has been investigated in generic noninteracting and interacting disordered systems, both critical and noncritical~\cite{DeChiara06, Znidaric2008, Bardarson2012, IgloiPRB12, Vosk2013, Pino14, Singh2015, Zhao16, DeTomasi2016}.
The XX spin chain (equivalent to the fermionic Anderson tight-binding model~\cite{Anderson1958}) with random external field (i.e. potential disorder) is an example of a noncritical  disordered system where all eigenstates are localized in the thermodynamic limit. The transverse-field Ising chain and the XX spin chain with bond-disorder are critical disordered one-dimensional (1D) systems which show a delocalization transition as a function of eigenstate energy~\cite{Eggarter78, Fisher95}.

A typical measure of entanglement in these systems is the entanglement entropy defined as the von-Neumann entropy of a subsystem for which we divide the 1D system into two equal parts (say $A$ and $B$) of length $L/2$. For a wavefunction $\left| \psi(t) \right \rangle$ of the total system, the von-Neumann entropy of subsystem A is given by
\bea
S(t) = -\mathrm{Tr}_A \rho_A(t) \ln \rho_A(t),\label{vonEE}
\eea
where $\rho_A(t)= \mathrm{Tr}_B \left| \psi(t) \right \rangle \left \langle \psi(t)\right|$ is the reduced density matrix of the subsystem $A$ at time $t$.

For the noncritical disordered noninteracting models of length $L$ when the localization length $\xi \gg L/2$ for weak disorder, the entanglement $S(t)$ grows linearly in time $t$ for $vt<L/2$ (where $v$ is a characteristic velocity~\cite{Calabrese2016}), 
and saturates to an extensive value showing a volume law for $vt \gg L/2$. Thus, the nature of entanglement dynamics in the weak disorder regime is similar to the clean case for small system sizes. For relatively stronger disorder when $\xi \lesssim L$, the initial growth of $S(t)$ until $vt \sim \xi$ roughly follows a power law with an exponent smaller than 1, and the saturated entanglement entropy at long times can be fitted as $S(L,t \to {\infty}) \equiv S_{\infty} \propto {\rm exp}(-L/2\xi)$ which allows one to obtain an estimate for $\xi$~\cite{Zhao16}. $S(t)$ becomes independent of $L$ when $\xi \ll L$ for strong disorder.

The dynamics of entanglement in critical disordered noninteracting models is quite nontrivial~\cite{IgloiPRB12,Zhao16}. In recent years a generalized real-space renormalization group (RSRG) approach including excited states has been applied to study $S(t)$ following a global quench in these models~\cite{Vosk2013}. It has been predicted from such studies of the critical XX spin chains in the thermodynamic limit that (a) the long-time growth of entanglement scales as $S(t) \sim \ln (\ln t)$, and (b) the saturated entanglement at long times scales as $S_{\infty} \sim \ln L$. High-precision numerical calculations~\cite{Zhao16} have observed an initial approximately logarithmic increase in $S(t)$ which is followed by $S(t) \sim \ln(\ln t)$ scaling over several orders of magnitude in time. However, these numerical studies have not found fully conclusive evidence for prediction (b). Earlier numerical studies in the critical random transverse-field Ising chain~\cite{IgloiPRB12} have found a regime where long-time entanglement $S(t) \approx 0.25~\ln (\ln t)$ + constant, and saturated  entanglement $S_{\infty} \approx 0.173~\ln L$ + constant, which are qualitatively consistent with the generalized RSRG predictions for the critical XX spin chains. 

In the presence of interactions, these 1D models are expected to show an MBL phase at higher disorder strengths, and the prediction (a) for the asymptotic growth of entanglement changes to $S(t) \sim \ln t$, while the prediction (b) for the saturated entanglement is a volume law, $S_{\infty} \sim L$~\cite{Bardarson2012,Vosk2013, Singh2015, Zhao16}. The dynamical entanglement entropy $S_{\infty}$ in the MBL phase is much smaller than that in the interacting delocalized (thermal) phase although both show a volume law. Some of the above features in the MBL phase have been observed in the XXZ spin chain with potential and bond disorders~\cite{Bardarson2012, Singh2015, Zhao16}. However, it is still not clear if there would be any difference in $S(t)$ due to the absence or presence of an underlying single-particle mobility edge~\cite{LiPRL15, Pino14}. The above results are based on many recent studies with short-range hopping and interactions in 1D disordered systems.

Although disordered models with long-range hopping and interactions are very relevant for experiments with ions~\cite{Smith2015}, dipolar molecules in an optical lattice and spin defects in a solid-state system~\cite{Yao2013d}, entanglement dynamics in such models has been relatively less investigated~\cite{Khatami2012,Pino14}. 
Another motivation for exploring long-range models is their relation to the higher-dimensional (D) ones.
The nature of MBL in higher-D short-range disordered models~\cite{Choi16, Chandran2016} is not yet clear as it is quite difficult to simulate reasonably large system size on classical computers. The long-range models in 1D can capture some features of the higher-D short-range models, such as the growth of the number of bonds with distance. It is thus interesting to see to what extent increasing long-rangeness in such models can be used to glean information on the effect of  increasing dimensionality  in short-range models.
In this work we consider spinless fermions on a 1D lattice of $L$ sites with long-range hopping and interactions. The Hamiltonian is given by~\cite{Burin2015}
\begin{equation} \label{HamiltonianLongRange}
H =- \sum_{i<j} J_{ij} (c_i^{\dg}c_{j} + c_{j}^{\dg}c_i) + \sum_{i<j} U_{ij} n_i n_j  - \sum_i \epsilon_i n_i, 
\end{equation}
where fermion occupation $n_i = c_i^\dagger c_i$, $J_{ij} = J / |i-j|^\alpha$, $U_{ij} = U / |i-j|^\beta$, $U>0$ for repulsive interactions and the $\epsilon_i$'s are chosen randomly from a uniform distribution $[-\eta,\eta]$.

We numerically study the dynamics of entanglement in the Hamiltonian in Eqn.~\ref{HamiltonianLongRange} for different values of $\alpha$ and $\beta$ in the absence and presence of interaction term $U_{ij}$. In the absence of interactions ($U_{ij}=0$), we observe that the scaling of long-time growth of entanglement and saturated entanglement shows interesting features for different $\alpha$ at relatively large disorder strengths. In particular, we find a special value of $\alpha$, $\alpha_{\rm log}$ when the long-time $S(t) \sim \ln t$ and $S_{\infty} \sim \ln L$. Also, the observed $L$-dependence of $S_\infty$ for smaller $\alpha$ is similar to that of the higher-D AL phase in short-range systems.

We then study the effect of long-range interactions along with long-range hopping on the growth of entanglement and the MBL transition in this system. For the strongly disordered interacting model, we find an intermediate regime in the window $1 \lesssim \alpha<3$ which shows a system-size dependent eigenstate entanglement and a strongly system-size dependent (likely a volume law) yet  small value of $S_{\infty}$. Using an analogy that smaller $\alpha$ corresponds to higher dimension in short-range systems, we expect a higher-D MBL phase in interacting short-range systems to show exactly the above two behaviors.
For $\alpha \gtrsim 3$ the effect of long-range hopping is quite reduced, and we recover the well-known properties of the (short-range) MBL phase in 1D such as slow-growth in long-time $S(t)$.

The paper is organized as follows.
In Sec.~\ref{model} we discuss the system parameters as well as the dynamical properties of the clean system, and provide some details of the simulations.
We study entanglement dynamics in the long-range noninteracting model (LRNM) in the presence of strong disorder in Sec.~\ref{FreeFermionQuench}.
We present results from the simulations of the full model with both long-range hopping and interactions in Sec.~\ref{LongRangeAndInteraction}. In Sec.~\ref{analogy} we explore the possibility of an analogy between 1D long-range models and higher-D short-range models and discuss properties of MBL phase of the latter from this perspective.
We summarize our results with a phase diagram in Sec.~\ref{Summary}.
We include four appendices: (\ref{App1})~to present single-particle eigenstate properties of the LRNM, (\ref{SectionXXZComparison}) to compare the entanglement dynamics in the LRNM to a short-range interacting model which displays an MBL phase, (\ref{App2})~to show the level-spacing and eigenstate thermalization in long-range interacting model and (\ref{App3})~to compare particle number fluctuations with $S(t)$ in the long-range interacting model.

\section{System and simulation details} \label{model}
For all physical systems, the exponents $\alpha$ and $\beta$ in Eqn.~\ref{HamiltonianLongRange} obey the constraint $\beta \le \alpha$, and if $\alpha = \beta$, $U \ge J$~\cite{Yao2013d, Gutman2015} where $J>0$ without loss of generality, and the lattice constant is set to unity. We also take $\alpha,\beta \gtrsim 1$ to ensure convergence of the energy density in the thermodynamic limit.
Some eigenstate properties of the above model in the noninteracting limit (i.e. $U= 0$) have been studied earlier~\cite{Rodriguez2003, BalagurovPRB2004}, and it has been found from these analytical and numerical investigations that the states at the top of the band are delocalized at moderate disorder strength ($\eta \sim J$) for $\alpha \le\alpha_c$ with $\alpha_c \approx 1.5$, and they undergo a delocalization-localization transition as the strength of disorder increases~\cite{Rodriguez2003}. All states of the noninteracting disordered chain are localized in the thermodynamic limit when $\alpha>\alpha_c$.

Following \textcite{Buyskikh2016}, the dynamical behavior of long-range models in the absence of disorder can be understood by analyzing the dispersion relation $\epsilon(k)$, maximum group velocity $v_g^{\rm max}$ and density of states in velocity $D_v(k)$ near $v_g^{\rm max}$. $D_v(k)$ is the number of states per volume with a velocity between $v_g$ and $v_g+\delta v_g$. The dispersion of LRNM without disorder is given by $\epsilon(k)=J\sum_{n\ne0}\f{e^{ikn}}{|n|^{\alpha}}$ which scales for small momenta as $\epsilon(k) \sim \epsilon_0+vk^{\alpha-1}$ for $1<\alpha<2$. In this regime $1<\alpha<2$, the dispersion is bounded while the $v_g^{\rm max} \sim \f{d\epsilon(k)}{dk} \sim k^{\alpha-2}$ is unbounded (infinite) for $k\to 0$. Also the density of states in velocity $D_v(k) \sim |\f{d^2\epsilon(k)}{dk^2}|^{-1} \sim k^{3-\alpha}$ is suppressed in this regime. As a result, even though correlations can build instantly via infinite $v_g^{\rm max}$ through the entire system, it overall grows slowly due to collective effect of combination  $v_gD_v(k) \sim k$ for $k\to 0$. 

For $\alpha>2$, both the $\epsilon(k)$ and $v_g^{\rm max}$ are bounded in clean LRNM but $D_v(k)$ near  $v_g^{\rm max}$ diverges. This implies that it is possible to excite infinitely many quasiparticles propagating with finite $v_g^{\rm max}$, which in turn build a well-defined front of entanglement for $\alpha>2$. We have confirmed the above features of entanglement dynamics in LRNM in the absence of disorder from our numerics. In this paper, we study how disorder influences the above description of entanglement growth by infinite and finite maximum group velocities for $\alpha<2$ and $\alpha>2$ in the absence and presence of long-range interactions.  

Here we mostly focus on the dynamical properties of the long-range system in Eqn.~\ref{HamiltonianLongRange}, preparing it in a simple initial state and evolving in time.
Such simple initial states can be considered as the ground state of an artificial Hamiltonian and this initialization is commonly known as a global quench~\cite{Bardarson2012}. We use a charge density wave at half-filling as the initial condition for all the results shown here. It is a state with one fermion at every other site, 
\bea
|\psi(t=0)\ra=\prod_{i=1}^{L/2} c_{2i}^{\dagger} |0\rangle,
\eea
where $|0\ra$ denotes the vacuum. We have also checked the robustness of our results by using random half-filled states as initial condition.
We show the results with open boundary conditions, but we have verified that they are qualitatively similar for periodic boundary conditions.
Even though we mostly present the results for entanglement entropy, we have also calculated Renyi entropies and particle number fluctuations $\mathcal{F}(t)$ in the system. $\mathcal{F}(t)$ is a measure of the quantum fluctuations in total number of particles in a subsystem and is given by
\bea
\mathcal{F}(t) = \left \langle \psi(t) \right| N_A^2 \left| \psi(t) \right \rangle -  \left \langle \psi(t) \right| N_A \left| \psi(t) \right \rangle^2,
\eea
where $N_A= \sum_{i\in A} c_i^\dagger c_i$ is total number of fermions in subsystem $A$. 
While Renyi entropies and $\mathcal{F}(t)$ behave qualitatively similar to the entanglement for the noninteracting system, interactions result in an important difference between the entropies and $\mathcal{F}(t)$, which we discuss in Sec.~\ref{LongRangeAndInteraction} and Appendix~\ref{SectionXXZComparison},\ref{App3}.

The time evolution for noninteracting fermions can be simulated using correlation matrices~\cite{Peschel2003}, and we can simulate reasonably long system sizes (upto $L=3200$ here) as the exact diagonalization is done for a single-particle problem.
For the noninteracting case we perform an averaging of $S(t)$, $\mathcal{F}(t)$, single-particle level-spacing ratio $\tilde{r}$ and entanglement in eigenstates $S_e$ over $10^4, 5000, 500$ realizations for $L\le400, L=800,1600, L=3200$ respectively, whereas for the interacting system this is done with $10^4, 10^3, 500$ realizations for $L\le12, L=14, L=16$ respectively.

\section{Global quench in the Long-range noninteracting model ($J \ne 0, ~U=0$)} \label{FreeFermionQuench}
\begin{figure}
\begin{center}
\includegraphics[width=0.99\linewidth]{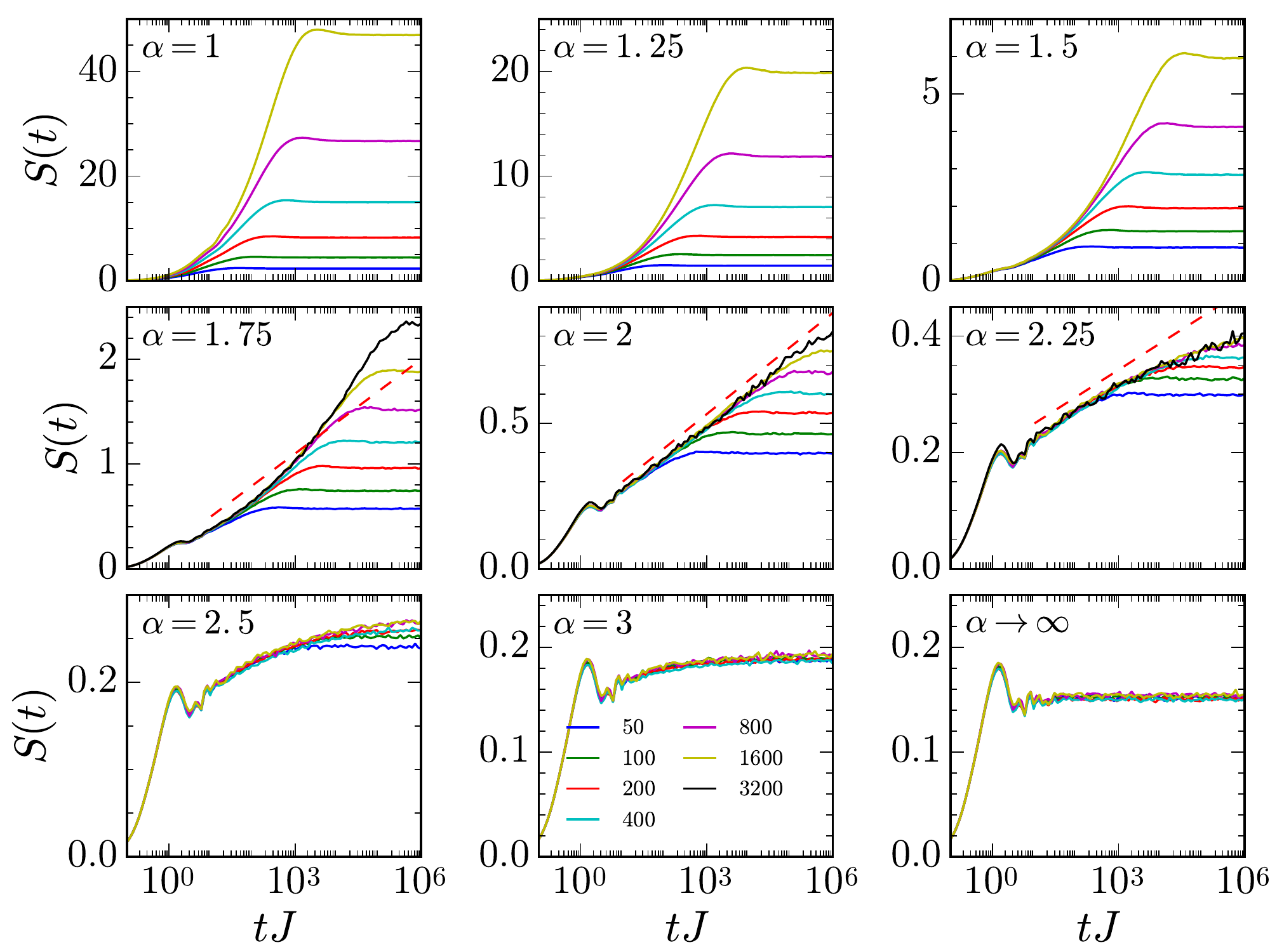}
\end{center}
\caption{(color online).
Time evolution of entanglement $S(t)$ in the long-range noninteracting model at different values of $\alpha$ for strong disorder $\eta \approx 20J$.
The chain lengths $L$ are $50,100,200,\cdots,1600$ in all except the middle row where $L=3200$ is also included.
The red dotted straight line in the middle row is a guide to eye for emphasizing the logarithmic growth at $\alpha=2$ in the center.
}
\label{timeEvolution}
\end{figure}
\begin{figure}
\begin{center}
\includegraphics[width=0.99\linewidth]{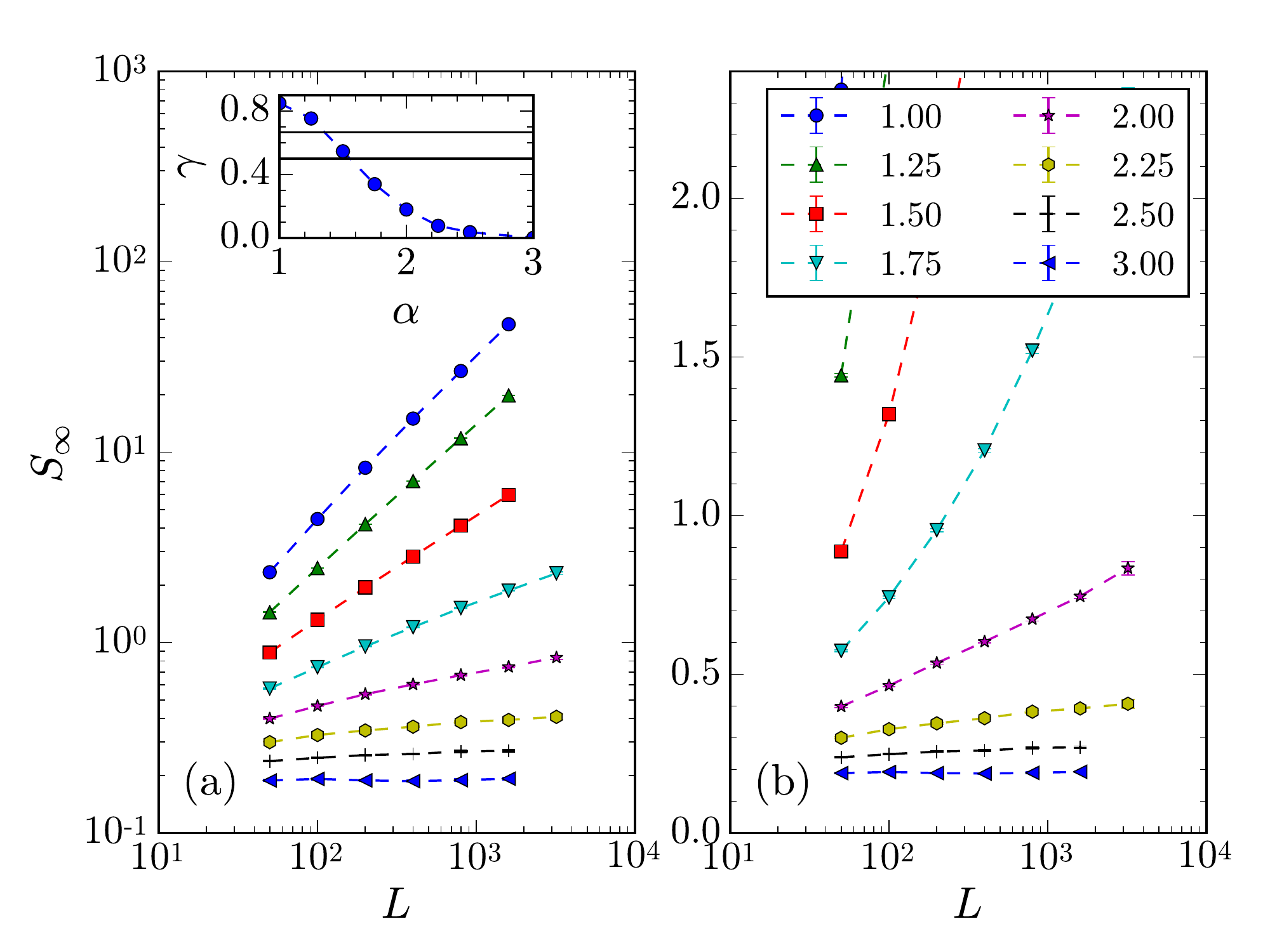}
\end{center}
\caption{(color online).
System-size $(L)$ dependence of asymptotic entanglement ($S_\infty$) in the long-range noninteracting model at $\eta \approx 20J$. $S_\infty$ shows an algebraic dependence on $L$ for smaller $\alpha$, as shown by a log-log plot in (a), while it grows as $\ln L$ at $\alpha=2$, highlighted by a log-linear plot in (b).
Different curves correspond to different values of $\alpha$.
In the inset of (a) we show the exponent $\gamma$ of algebraic $L$-dependence ($S_\infty \sim L^\gamma$) as a function of $\alpha$.
The two solid black lines in the inset represent the $\gamma$ values corresponding to an area law in 2D ($1/2$) and 3D ($2/3$) short-range models.
}
\label{SystemSizeDependence}
\end{figure}

We first summarize the single-particle eigenstate properties of the disordered LRNM from our studies of level-statistics and entanglement (see Appendix~\ref{App1}). We confirm the presence of delocalized states at the top of the band at moderate disorder for $\alpha \le \alpha_c$ with $\alpha_c \approx 1.5$, and a delocalization-localization transition with increasing $\eta$ as predicted earlier in \textcite{Rodriguez2003} and \textcite{BalagurovPRB2004}. Other states in the band of a long chain are localized even at relatively low disorder for $\alpha \le \alpha_c$. On the other hand, all states of the noninteracting chain for $\alpha > \alpha_c$ are localized at any disorder.

We now discuss the entanglement dynamics in the LRNM at a strong disorder ($\eta \approx 20J$) when all the single-particle eigenstates, including those at the top of the band, are localized.
In Fig.~\ref{timeEvolution} we show $S(t)$ for different $\alpha$ and chain lengths. For smaller $\alpha$, $S(t)$ shows a rapid growth followed by a $L$-dependent saturation (first row of Fig.~\ref{timeEvolution}). $S(t)$ starts to display a slow logarithmic growth after a short rapid initial growth for $\alpha>\alpha_c$.
Around $\alpha=\alpha_{\mathrm{log}}=2$, $S(t)$ grows logarithmically for the longest time before saturating due to finite-size effects. We highlight this very interesting feature  at $\alpha=2$ by comparing it to the growth at nearby values of $\alpha$ (=$1.75$ and $2.25$) in the middle row of Fig.~\ref{timeEvolution}. Although $S(t)$ at all three $\alpha$ appears to show a logarithmic growth in the beginning, later it grows faster at $\alpha=1.75$ while it does so more slowly at $\alpha=2.25$. The longest duration of logarithmic growth at $\alpha=2$ can be qualitatively seen by comparing the time evolution to the red (dashed) straight line in each subplot. $\alpha=2$ is also special because only here does the saturated entanglement scale as $S_{\infty} \sim \ln L$ (see below). Though the slow logarithmic growth of $S(t)$ at $\alpha_{\rm log}$ in the LRNM is somewhat similar to that in the MBL phase of a random XXZ chain with nearest-neighbor interactions, there are some subtle differences in the entanglement dynamics between these two models which we discuss in  Appendix~\ref{SectionXXZComparison}.

For $\alpha \gtrsim 3$, $S(t)$ becomes independent of $L$, and only shows a short early growth due to the movement of particles within the localization length. The limit $\alpha \to \infty$ implies a short-range model, which represents a 1D XX spin chain with potential disorder (non-critical), and it shows an AL phase at any disorder strength in the thermodynamic limit.

In Fig.~\ref{SystemSizeDependence} we show the $L$-dependence of $S_{\infty}$ for different values of $\alpha$.
For $\alpha<\alpha_{\rm log}$, $S_{\infty}$ shows an algebraic dependence on $L$ (seen as straight lines on log-log plot), while $S_{\infty}$ becomes independent of $L$ for $\alpha \gtrsim 3$. $S_{\infty}$ scales as $\ln L$ at $\alpha_{\rm log}$.  
We show the exponent $\gamma$ of the algebraic $L$-dependence (i.e., $S_\infty \sim L^\gamma$) as a function of $\alpha$ in the inset of Fig.~\ref{SystemSizeDependence}~(a).
The exponent $\gamma$ falls with increasing $\alpha$ and is less than $1$ in the window $1 \lesssim \alpha < 3$.
We also show there the values of $\gamma$ corresponding to an area law in 2D and 3D short-range  models by solid black lines.
On a log-log plot, a logarithmic dependence can of course appear algebraic with a small exponent if the x-range is not sufficiently long. Fig.~\ref{SystemSizeDependence}~(b) shows the same data on a log-linear scale to emphasize the quality of the log behavior at $\alpha=2$.

The initial rapid growth in this system for all allowed $\alpha$ is due to propagation of quasiparticles within the localization length which falls with increasing disorder. The logarithmic growth of long-time $S(t)$ at $\alpha=2$ can be a reminiscence of the feature of underlying maximum group velocity $v_g^{\rm max}$ in clean systems which changes from infinite to finite across $\alpha=2$. The exponent $\gamma_t$ (long-time $S(t) \sim t^{\gamma_t}$) for smaller $\alpha~(<2)$ seems to suggest a drifting exponent of $\gamma_t$ which goes through zero at $\alpha_{\rm log}$ in response to increasing $\alpha$. Therefore, we can perceive that the appearance of logarithmic growth at $\alpha_{\rm log}$ is a consequence of culmination of power-law growth of long-time $S(t)$ there, which is again due to a drifting power-law dependence of some underlying physical quantity like  $v_g^{\rm max}$. In our numerics we have seen some slower than logarithmic growth of long-time $S(t)$ for $\alpha>2$. However, it is difficult to precisely determine what is the nature of such growth as well as at which value of  $\alpha$ such growth completely stops in the thermodynamic limit. 

The observed $L$-dependence of $S_\infty$ in the window $1\lesssim \alpha <3$ is surprising as, in contrast, we know from the single-particle eigenstate properties at such strong disorder that all states are fully Anderson localized for all $\alpha \gtrsim 1$. However, even the entanglement in many-particle eigenstates shows a $L$-dependence similar to $S_\infty$. Because $S_\infty$ and  many-particle eigenstate entanglement in the LRNM show a $L$-dependence in the window $1 \lesssim \alpha <3$ while they both are $L$-independent for $\alpha \gtrsim 3$, we may call the former regime a {\it quasi-AL} while the latter one a true AL. 

\section{Long-range interacting model} \label{LongRangeAndInteraction}
\begin{figure*}
\includegraphics[width=0.99\linewidth]{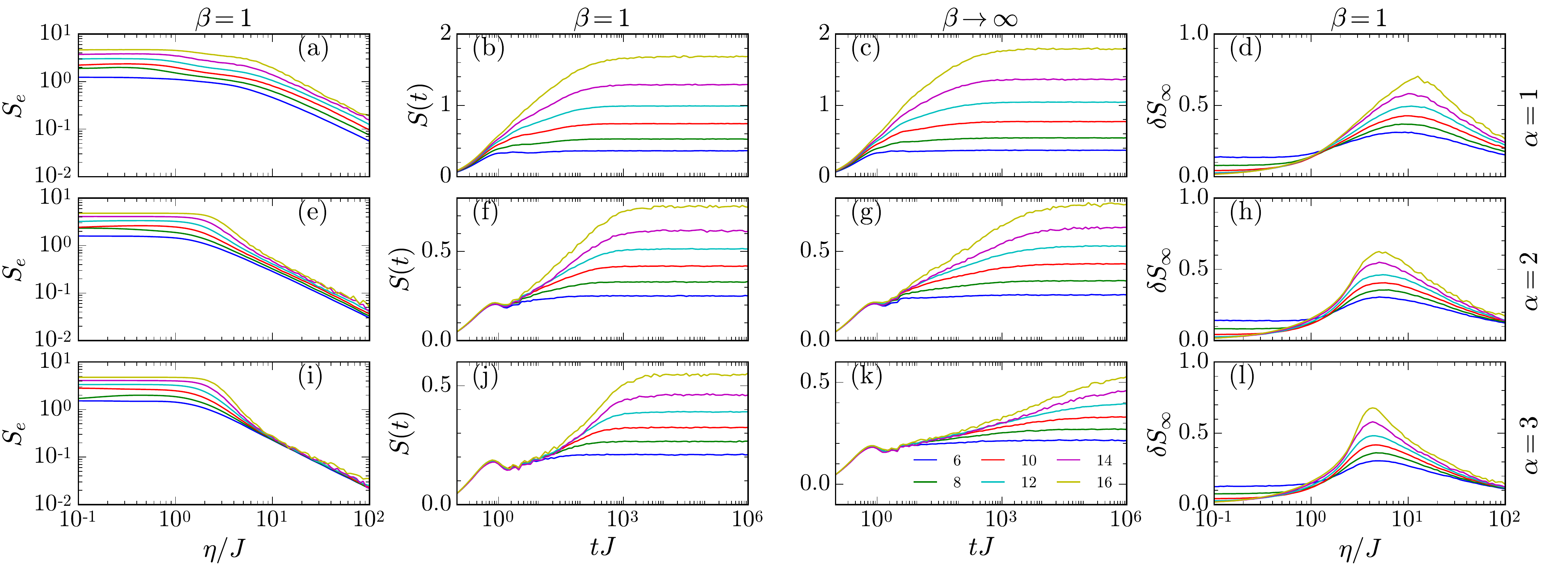}
\caption{(color online).
Eigenstate and dynamical entanglement properties in long-range interacting model at different values of $\alpha~(=1,2,3)$ (rows).
The first column shows the eigenstate entanglement $S_e$ in the middle of the spectrum as a function of disorder ($\eta$) for $\beta=1$.
Time evolution of entanglement $S(t)$ for $\beta~(=1,\infty)$ (second and third columns respectively) for $\eta \approx 20J$.
The last column shows the fluctuations in asymptotic dynamical entanglement $\delta S_\infty$ with disorder $\eta$ at three $\alpha$ values for $\beta=1$.
}
\label{timeEvolutionInteracting}
\end{figure*}

We now discuss the properties, both of eigenstates and dynamical entanglement, of the long-range interacting model in Eqn.~\ref{HamiltonianLongRange} for general values of $\alpha, \beta$.
For low disorder, we find the system in an extended phase for all values of $\alpha$ and $\beta$, as expected. This can be inferred from the extensive and large entanglement in the eigenstate as seen in the first column of Fig.~\ref{timeEvolutionInteracting} at low disorder.
However, the behavior of entanglement for strong disorder has an intricate dependence on these parameters, and we discuss it here in some detail.
For simplicity, we start with the situation closest to the well-studied case of short-range systems, namely large values of $\alpha$ and $\beta$.
For $\alpha=3$ and $\beta \rightarrow \infty$ we find the usual logarithmic growth of entanglement at high disorder (Fig.~\ref{timeEvolutionInteracting}~(k)), while changing $\beta$ to $1$ changes this slow growth from logarithmic to algebraic (Fig.~\ref{timeEvolutionInteracting}~(j))~\cite{Pino14}. Though the nature of growth (logarithmic vs. algebraic) is not obvious immediately in Fig.~\ref{timeEvolutionInteracting}~(k,j), it becomes clear when the same $S(t)$ data is plotted with different scale for y-axis (log-linear vs log-log).
The nature of the slow growth at $\alpha \gtrsim 3$ is mostly independent of $\alpha$ and depends strongly on $\beta$.
The slow growth is algebraic in time for the long-range interactions (small $\beta$) and is logarithmic in time for short-range interactions (large $\beta$).
This is in agreement with earlier studies on the role of long-range interactions in a short-range hopping model~\cite{Pino14}.
However, this dynamical difference in $S(t)$ is not present at the level of eigenstate entanglement which shows an area law for all values of $\beta$.
Irrespective of the value of $\beta$ when $\alpha \gtrsim 3$, the system is in the MBL phase at high disorder, characterized by a logarithmic (algebraic) growth  of $S(t)$ in presence of short (long)-range interactions and an area law for entanglement in all eigenstates.
For $1 \lesssim \alpha<3$ and high disorder, the slow growth of $S(t)$ at later time, if present, is a very short one in contrast to the short-range MBL phase (Fig.~\ref{timeEvolutionInteracting}(f,g)).
However, the characteristic of dynamical entanglement in this region is also different from the delocalized case in that $S_{\infty}$ here is much smaller than the latter.
The difference of this region from the extended phase can also be seen in the eigenstate entanglement $S_e$ which is again much smaller (compare $S_e$ between low and high disorder in Fig.~\ref{timeEvolutionInteracting}(a,e)).
Nevertheless, unlike the MBL phase at $\alpha \gtrsim 3$ this small entanglement in the eigenstates has a clear system-size dependence.
 Since the strongly disordered regime in the window $1 \lesssim \alpha<3$ appears to share some properties with the (short-range) MBL phase (such as small entanglement), but not others (including a long slow-growth of $S(t)$ and an area law for eigenstate entanglement), we call this regime {\em quasi-MBL}. 
From a dynamical point of view, the entanglement growth in this phase has a significant contribution due to movement of particles along with the dephasing mechanism.
This result can be inferred from the observation that the particle number fluctuations $\mathcal{F}(t)$, which are insensitive to the dephasing mechanism, behave similar to the dynamical entanglement in this phase, in contrast to the usual short-range MBL phase (see Fig.~\ref{SystemSizeDependenceInteracting}).

Both the asymptotic dynamical entanglement $S_{\infty}$ and eigenstate entanglement $S_e$ (see e.g. first column of Fig.~\ref{timeEvolutionInteracting}) fall from a constant ($\eta$-independent) and large value to some small and $\eta$-dependent values with increasing $\eta$ for all $\alpha \gtrsim 1$. This change in behavior from large to small $S_{\infty}$ or $S_e$ with increasing $\eta$ is related to a delocalization-localization transition separating the interacting delocalized phase at weak $\eta$ and localized (MBL) phase at large $\eta$.
We can try to locate the MBL transition by considering the standard deviation of $S_\infty$ with respect to disorder namely $\delta S_{\infty}$~\cite{Singh2015}. We present it in the last column of Fig.~\ref{timeEvolutionInteracting}, and find diverging $\delta S_{\infty}$ near the transition in all the long-range scenarios.

We find that both the asymptotic properties as well as the eigenstate properties are somewhat insensitive to $\beta$, so we show the results for $\beta=1$ only.
Thus, the role of long-range interactions is mostly to speed up the later growth in $S(t)$. On the other hand, the long-range hopping plays a role in the initial rapid growth of $S(t)$ for $\alpha < 2$ which is $L$-dependent from the start  implying an infinite $v_g^{\rm max}$.
The effect of $\alpha$ on the entanglement in the eigenstates at high disorder is a shift from a volume law to an area law with an increasing $\alpha$, while the value of $S_e$ being significantly smaller than the delocalized phase in all cases (first column in Fig.~\ref{timeEvolutionInteracting}).
As the role of long-range hopping becomes significant with smaller $\alpha$, the transition shifts to larger disorder.

\section{1D long-range  and higher-dimensional short-range models}\label{analogy}
Some results of the long-range system suggest attempting an analogy to higher-D short-range ones in order to compare
and contrast the two types of behaviors.
The analogy in two limiting cases is straightforward.
The limit $\alpha \to 0$ in the 1D LRNM corresponds to an  infinite-range hopping system, i.e. a fully connected model, and is thus
akin to a certain type of infinite-dimensional noninteracting system.
By contrast,  the limit of $\alpha \to \infty$ in the 1D LRNM is equivalent to a 1D short-range noninteracting model as in this limit every site is effectively connected only to two of its nearest neighbors, as the ratio of next-nearest to nearest neighbor coupling vanishes.

There are many ways of defining effective dimensionality, the most suitable one depends on the question being asked. For short-range 
models, these usually agree, e.g.\ a common definition involving
 the scaling of number of sites with linear size $L$. For our 1D LRNM, this gives a value of $1$ independent of $\alpha$, as it is the 
 coupling strength at long distances which varies with $\alpha$, not the number of sites per se. 

For our purposes, a practically useful and consistent definition is the following: we construct a simple bipartition of the system into two equal parts, and ask about the total strength of the bonds of the hopping graph that minimally needs to be cut. 
For the 1D LRNM, such a bipartition cuts a single bond between nearest neighbor sites, two bonds between next-nearest neighbor sites and so on at each cut (1D system with periodic boundary has two cuts). Thus, we find the total effective hopping strength:
$$J_{\rm eff} = 2 \times \frac{J}{1^\alpha} + 4 \times \frac{J}{2^\alpha} + 6 \times \frac{J}{3^\alpha} + \ldots
\approx 2J\sum_{n=1}^{L/2}|n|^{1-\alpha}$$
across a bipartition boundary in the middle of the 1D LRNM. The summation is an approximation as we have double counted the $n=L/2$ term. In deriving $J_{\rm eff}$ we take the contribution of a long-range bond to be independent of the location of the cut, i.e. a bond that is cut symmetrically with equal half on each side has the same contribution as ones (of same length) which are cut asymmetrically. This is an approximation for bonds beyond next-nearest neighbors.
From the expression of $J_{\rm eff}$ we find that total effective hopping strength across a boundary diverges with system size $L$ for $\alpha \le 2$. For example, we find $J_{\rm eff} \sim L$ for $\alpha=1$ and $J_{\rm eff} \sim {\rm ln}L$ for $\alpha=2$ when $L \to \infty$.

The corresponding calculation of total hopping strength $J^d_{\rm eff}$ across a $(d-1)$-dimensional surface/boundary in the middle of a $d$-dimensional noninteracting model with short-range hopping gives an asymptotic dependence of $J^d_{\rm eff} \sim L^{(d-1)}$ where $L$ is the length scale of the $d$-dimensional short-range model.
In terms of the volume $V \sim L^d$ of the short-range model, the total hopping strength across the boundary becomes $J^d_{\rm eff} \sim V^{(d-1)/d}$.
This asymptotic dependence is independent of specific lattice structure  of the short-range model. For the 1D LRNM, the volume $V \sim L$. Thus, when we compare the asymptotic $V$-dependence of $J_{\rm eff}$ of 1D LRNM with that of $J^d_{\rm eff}$ of $d$-dimensional short-range model, we find $\alpha \approx 1.5, 1.34$ of the 1D LRNM respectively relates to $d=2,3$ dimensional noninteracting short-range model.

We now discuss some of our results in light of this analogy.
The  quasi-AL phase in Sec.~\ref{FreeFermionQuench} is akin to an AL phase in a higher-D short-range noninteracting model: the entanglement in many-particle eigenstates as well as $S_{\infty}$ show area-law in the AL phase of a short-range noninteracting model of any dimension and we have confirmed this for the short-range model numerically. Interestingly, the (interpolated) values of $\alpha \approx 1.56, 1.36$ showing respectively $\gamma=1/2, 2/3$ for the system size-dependence (satisfying area-law of entanglement) of a 2D and 3D short-range noninteracting model in the AL phase are very close to the predicted values $1.5, 1.34$ from the above analogy.
We can also argue that the asymptotic $L$-dependence of $S_{\infty}$ and entanglement in many-particle eigenstates at $\alpha=1,2$ is due to that of $J_{\rm eff}$ at these $\alpha$'s respectively.

Even though these eigenstate and asymptotic dynamical properties agree, the analogy is limited in that the time-scale involved in the global quench can be very different between 1D LRNM and higher-D short-range model. This is because, in the higher-D short-range model, the particles move relatively smaller distances (typically a localization length) over a larger boundary; thus asymptotic dynamical properties are achieved in a shorter time.
On the other hand in the 1D LRNM, there are many weak hoppings across a point boundary whose contribution in the asymptotic dynamical properties takes a very long time to appear (hopping term of strength $J/R^\alpha$ takes $\sim R^\alpha/J$ time to saturate).

Assuming a similar analogy between the 1D long-range model and the higher-D short-range model exists in the presence of interactions, we now reconsider the following quantities: $S_e$, many-particle level-spacing ratio $\tilde{r}$ and eigenstate thermalization. Though it is hard to find the exact $L$-dependence of $S_e$ from our simulations with limited system sizes, we do see some signatures of an area law of $S_e$ for the interacting system in a higher dimension (2D and 3D correspond respectively to $\alpha \approx 1.56, 1.36$).
We note here that $S_e$ and other properties discussed below do not depend on the value of $\beta$, and the analogy between 1D long-range system and a higher-D short-range system is mainly due to $\alpha$.
The long-range model near $\alpha=1$ for which we find a signature of an almost extensive $S_e$  in Fig.~\ref{timeEvolutionInteracting}(a) is also consistent with such an analogy as small $\alpha$ corresponds to a very high $d$ dimensional short-range model, for which there would be a little difference between volume and area law of entanglement ($\gamma$ being $1$ and $1-1/d$ respectively). 

We also observe Poisson level spacing and absence of eigenstate thermalization for $1<\alpha<3$ in the long-range interacting model (see Appendix~\ref{App2}) at large disorder suggesting a similarity in the eigenstate properties of MBL phase in 1D and higher-D short-range models by our analogy. Therefore, our analogy would suggest the existence of an MBL phase in higher-D short-range interacting models~\cite{Basko2006, Choi16, Chandran2016}, which has area-law for $S_e$, Poisson level spacing and  absence of eigenstate thermalization. Our quasi-MBL phase in Sec.~\ref{LongRangeAndInteraction} would then be related to the MBL phase in higher-D short-range interacting models.
Just like the noninteracting case we expect here some qualitative difference in the dynamics between higher-D short-range interacting model and the corresponding 1D long-range one.
The long time growth of entanglement in the former will be exclusively due to the dephasing mechanism whereas for the latter there will also be a contribution due to long-range particle hopping.

\section{Summary} \label{Summary}
\begin{figure}
\begin{center}
\includegraphics[width=0.99\linewidth]{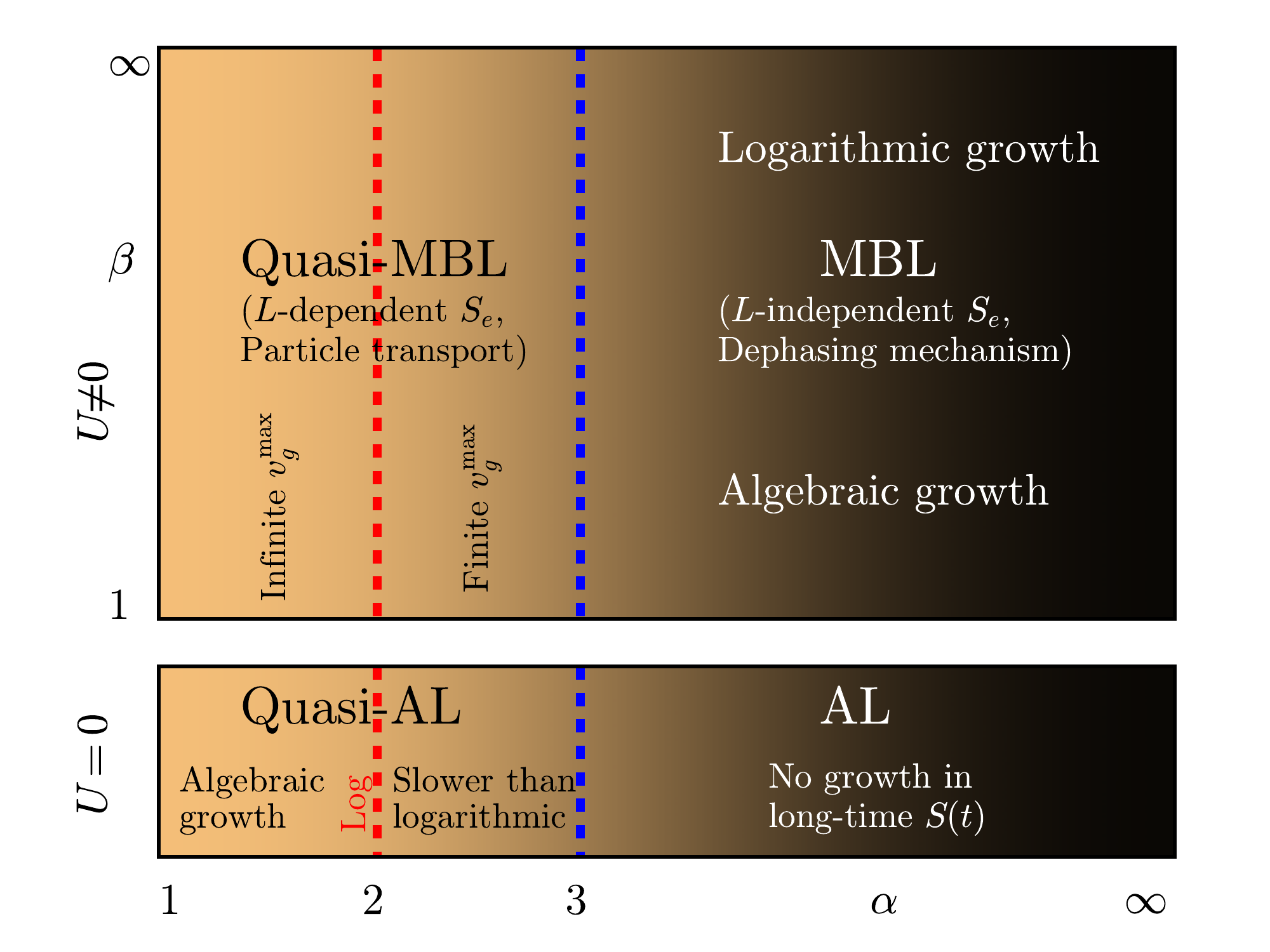}
\end{center}
\caption{(color online).
A cartoon phase diagram of the spinless fermion model in 1D with long-range hopping at strong disorder without (bottom panel) and with (top panel) interactions  decaying with distance $r$ as $U/r^{\beta}$. The dashed blue line at $\alpha \approx 3$ divides the long-range noninteracting model in quasi-AL (with some growth of long-time entanglement, $S(t)$) and AL (without any  growth of long-time $S(t)$) phases, and it divides the long-range interacting model in  quasi-MBL (size-dependent eigenstate entanglement $S_{e}$ and particle transport during long-time dynamics) and MBL (size-independent $S_{e}$ and the dephasing mechanism) phases.  The dashed red line at $\alpha \approx 2$ points the region of logarithmic growth of long-time entanglement in the long-range noninteracting model. The maximum group velocity $v_g^{\rm max}$ of quasi-particles in the long-range (noninteracting and interacting) model is unbounded for $\alpha<2$ and bounded for $\alpha>2$. The growth of long-time entanglement in the MBL phase is algebraic for small $\beta$ and logarithmic for large $\beta$.   
}
\label{cartoon}
\end{figure}

To summarize, we have studied entanglement dynamics and MBL in a critical disordered 1D model with long-range hopping and interactions. One specialty of this model is that we could separately investigate the roles of long-range hopping and interactions for the entanglement growth and MBL transition. It is not possible to separate these contributions in many long-range spin models, e.g., XY~\cite{Burin2015}, transverse-field Ising~\cite{Smith2015} and XXZ models~\cite{Yao2013d} which have been examined recently in relevant contexts.

In recent years, a slow logarithmic growth of $S(t)$ at large $t$ after a quench has been associated with the MBL phase to separate it from the related Anderson insulator, which does not show such logarithmic growth in previously studied (short-range) disordered models. Here we have found a slow  logarithmic growth of  $S(t)$ in a highly disordered LRNM at $\alpha=2$. The eigenstate properties of the LRNM for all allowed $\alpha$ clearly signal that all states are Anderson localized at such high disorder. 

Thus, our results for the LRNM reveal that the logarithmic growth of $S(t)$ is not a sufficient condition for separating the MBL phase from the AL phase in all disordered models.
The quasi-MBL phase at smaller $\alpha$ or the MBL in higher-D short-range models seem to have  the same eigenstate properties of the MBL phase of 1D short-range models, namely Poisson level spacing and absence of eigenstate thermalization~\cite{Chandran2016}. However, the quasi-AL phase at smaller $\alpha$ or the AL in higher-D short-range models also share these eigenstate properties. This needs to be borne in mind when  aiming to unambiguously detect the MBL phase in experiments in 1D long-range systems. The slow growth of entanglement due to particle transport in the LRNM can be mistaken for that due to dephasing mechanism in the MBL phase. It may thus be necessary to do a more quantitative analysis of, e.g., 
the difference between time-scales of such slow growth due to different mechanisms, in order to experimentally distinguish between the MBL and AL phases.

The growth of $S(t)$ at long time in the LRNM for $1 \lesssim \alpha<3$ results in an $L$-dependent saturated entanglement even at very large disorder, which we call quasi-AL (see the phase diagram in Fig.~\ref{cartoon}).
Similarly, we have a quasi-MBL phase at large disorder in the presence of interactions in this model for $1 \lesssim \alpha<3$ which is related to the higher-D MBL. The entanglement dynamics in the noninteracting and interacting long-range model for $\alpha \gtrsim 3$ at large disorder shows commonly accepted features of the AL (e.g., area law for $S_{\infty}$) and MBL (e.g., slow growth in $S(t)$) phases in 1D short-range models. We conclude that quasi-AL and AL which occur respectively for $1 \lesssim \alpha<3$ and $\alpha \gtrsim 3$ in the LRNM, turn into quasi-MBL and  MBL  as interactions are turned on. 
 However, we should emphasize that the boundaries between different phases are somewhat crude due to limitations of numerical calculations. 

Finally, while the existence of MBL and its properties in higher-D models are not yet settled, our numerical calculations with 1D long-range system provide some concrete pointers towards the physics of  MBL phase in 
higher-D interacting models.
Additionally, our analogy would suggest that such an MBL phase in higher-D short-range systems could have some accepted properties of 1D MBL phase, such as breakdown of eigenstate thermalization, Poisson level statistics and an area law for eigenstate entanglement. On the other hand, recent work of \textcite{Chandran2016} using an extended l-bit model proposes that the higher-D MBL phase will show many eigenstate properties similar to the thermal state (delocalized phase).

While we have seen good agreement of our analogy for noninteracting fermions in our numerics,  it is not obvious 
whether an analogy centering on the parameter $\alpha$ works in the presence of interactions. Also, the role of $\beta$ in this analogy is not settled, and could in principle 
lead to a further set of distinctive behaviors; here, however, we have numerically observed that both the eigenstate and asymptotic properties hardly depend on $\beta$, suggesting as the simplest scenario that such an analogy, if it works for interacting systems similarly well, would be independent of $\beta$. 

In either case, existence and nature of  a presumptive  MBL phase in higher-D  requires further study for a definitive answer. Conceivably, given the limits of numerical approaches in higher-D,  experiments will be helpful in settling this issue---there is already some experimental evidence supporting the existence of MBL phase and breakdown of thermalization in 2D \cite{Choi16}.

\section*{Acknowledgments}
We thank A. Pal, M. Haque and A. M. Garc{\'{i}}a-Garc{\'{i}}a for discussions. This work was in part supported by DFG via SFB 1143. RS thanks the Israel Council for Higher Education's Planning and Budgeting Committee (CHE/PBC) and Israel Science Foundation's grants N. 1452/14 and N. 231/14 for financial support.
DR acknowledges funding from the Department of Science and Technology, India via the Ramanujan Fellowship.

\appendix
\setcounter{figure}{0}
\renewcommand\thefigure{A\arabic{figure}}

\section{Eigenstate properties of the long-range noninteracting model ($J \ne 0, ~U=0$)}
\label{App1}
\begin{figure}
\begin{center}
\includegraphics[width=0.99\linewidth]{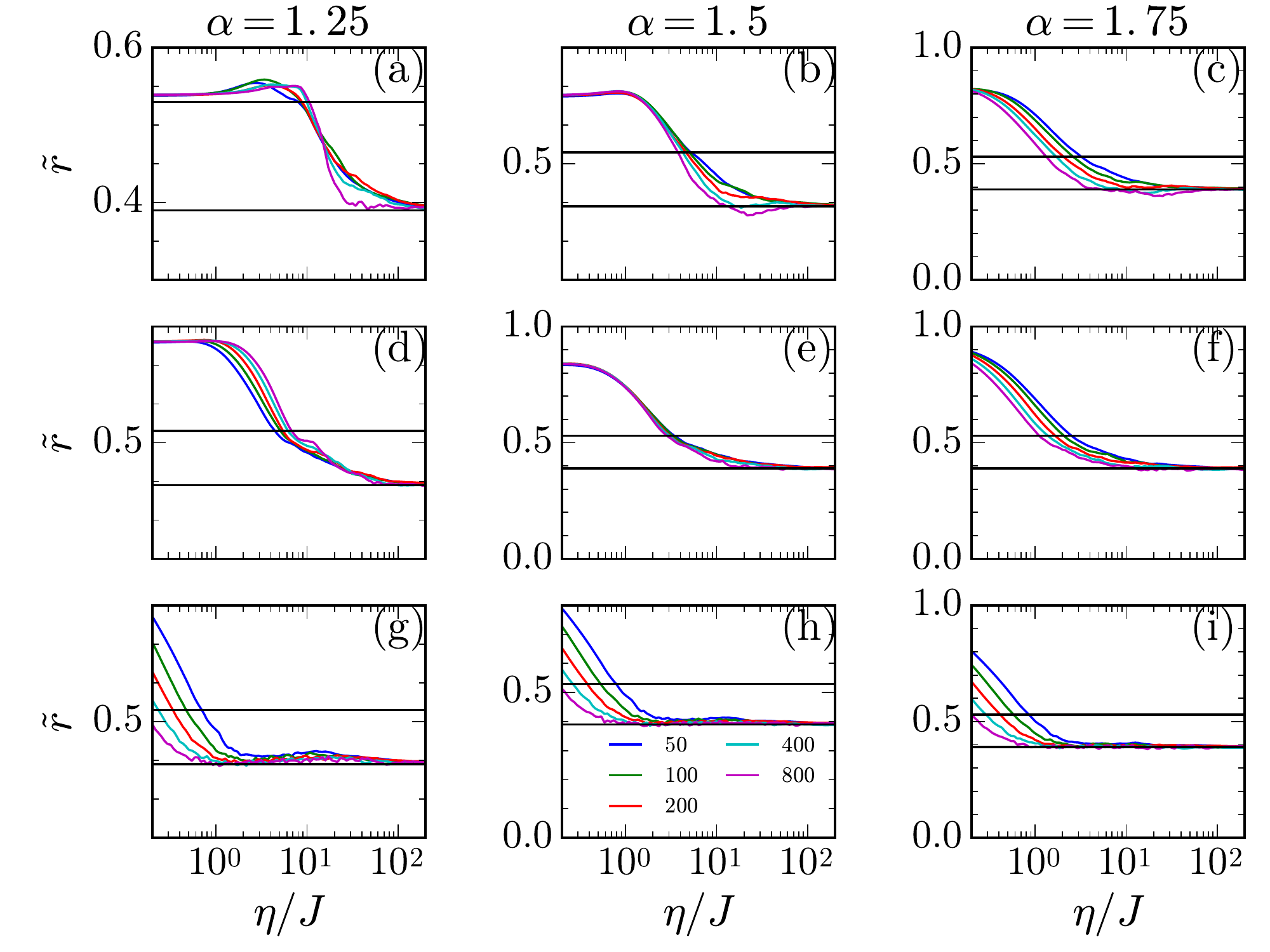}
\end{center}
\caption{(color online).
Single-particle level-spacing ratio ($\tilde{r}$) in the long-range noninteracting model for the highest energy eigenstate (top row), the second highest energy eigenstate (middle row) and the eigenstate in the middle of the spectrum (bottom row) for three values of $\alpha$ ($1.25, 1.5, 1.75$). 
The two solid lines in each subplot show the GOE ($\approx 0.536$) and the Poisson ($\approx 0.386$) values for $\tilde{r}$.
}
\label{singleParticleLevelSpacing}
\end{figure}
\begin{figure}
\begin{center}
\includegraphics[width=0.99\linewidth]{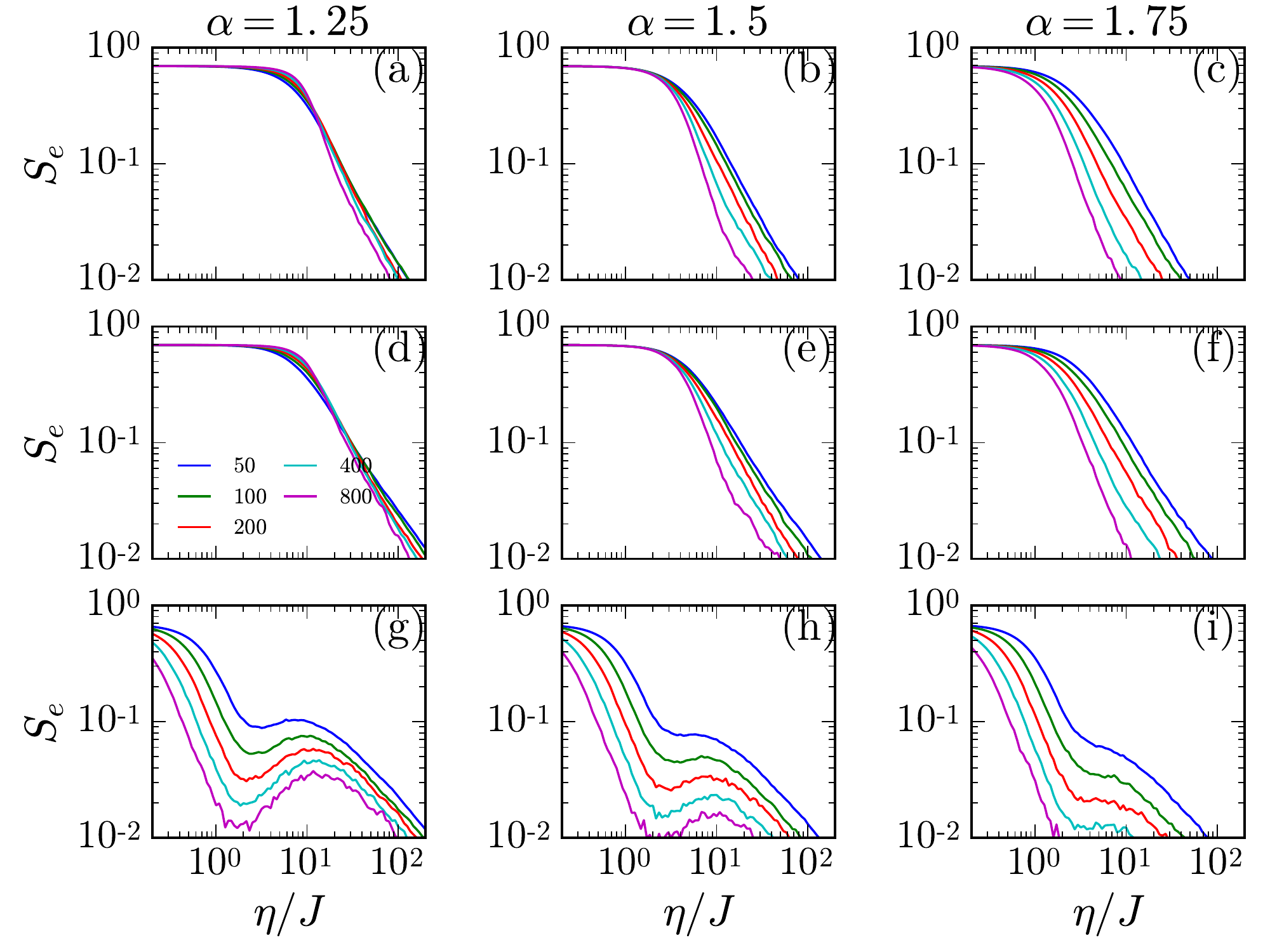}
\end{center}
\caption{(color online).
Single-particle eigenstate entanglement $S_e$ in the long-range noninteracting model in the highest energy eigenstate (top row), the second highest energy eigenstate (middle row) and the eigenstate in the middle (bottom row) for three values of $\alpha$ ($1.25, 1.5, 1.75$). 
}
\label{singleParticleEntanglement}
\end{figure}

We investigate the properties of single-particle eigenstates of the LRNM ($J \ne 0, ~U=0$) with increasing disorder strength at different $\alpha$. In Fig.~\ref{singleParticleLevelSpacing} we show the level-spacing ratio $\tilde{r}$ between different eigenstates following a definition by \textcite{Oganesyan2007}
$$\tilde{r}_i = \mathrm{min} \left(r_i, \frac{1}{r_i} \right), r_i = \frac{\epsilon_{i+1}-\epsilon_i}{\epsilon_i-\epsilon_{i-1}}$$
where $\epsilon_i$'s are the single-particle eigenvalues. This quantity has turned out to be a good discriminator for different universal level-spacing statistics. The value of $\tilde{r}_i$ is $4-2\sqrt{3} \approx 0.536$ for the Gaussian orthogonal ensemble (GOE) indicating extended states, while it is $2\ln2-1 \approx 0.386$ for the Poisson level spacing if the states are localized. We show these two values as solid black lines in each subplot of Fig.~\ref{singleParticleLevelSpacing}.
The top row of Fig.~\ref{singleParticleLevelSpacing} shows $\tilde{r}_i$ at the top of the spectrum (calculated using top three eigenstates with highest energy, $i=L-1$), while the middle row is calculated using three highest energy eigenstates excluding the top ($i=L-2$). The bottom row shows the data for the eigenstate in the middle ($i=L/2$). As expected we find that the chain shows a Poisson level spacing at very high disorder in all the plots. We can also identify that the departure from Poisson statistics is only a finite-size effect for the middle of the spectrum (bottom row) as well as for $\alpha=1.75$ (last column), since all these curves tend towards the Poisson value upon increasing the system size systematically. The characteristics of the remaining four plots (first two rows and first two columns) are more complicated as the departure from a Poisson behavior does not appear to be a finite-size effect and only one of them is close to the GOE value. The eigenstate at the top of the band for $\alpha \lesssim 1.5$ is expected to be extended~\cite{Rodriguez2003, BalagurovPRB2004}, and the level-spacing for this case shown in Fig.~\ref{singleParticleLevelSpacing}~(a) is consistent with that. However, the level-spacing ratio for next eigenstate shows a departure from the GOE behavior even for $\alpha \lesssim 1.5$.

The entanglement $S_e$ in the eigenstates of the LRNM also shows features of a delocalization-localization transition near the top of the band.
For larger disorder we get very small entanglement in all cases, while we get the maximum allowed entanglement ($\ln 2$) at very low disorder for states near the top of the band (see Fig.~\ref{singleParticleEntanglement}~(a-f)).
Similar to the level-spacing, the finite entanglement for low disorder at the middle of the spectrum (bottom row) can be perceived to be a finite-size effect.
Both the highest and second highest eigenstate of the spectrum for $\alpha=1.25$ show a transition at a finite $\eta$.
The behavior for higher $\alpha$ near the band-top is again more complicated and the data presented here is not conclusive, but one can still notice some finite-size effects which suggest that there is no transition.

\section{Long-range noninteracting model ($J \ne 0, ~U=0$) vs. short-range interacting model ($\alpha,\beta \to \infty, ~J,U \ne 0$)} \label{SectionXXZComparison}
\begin{figure}
\begin{center}
\includegraphics[width=0.99\linewidth]{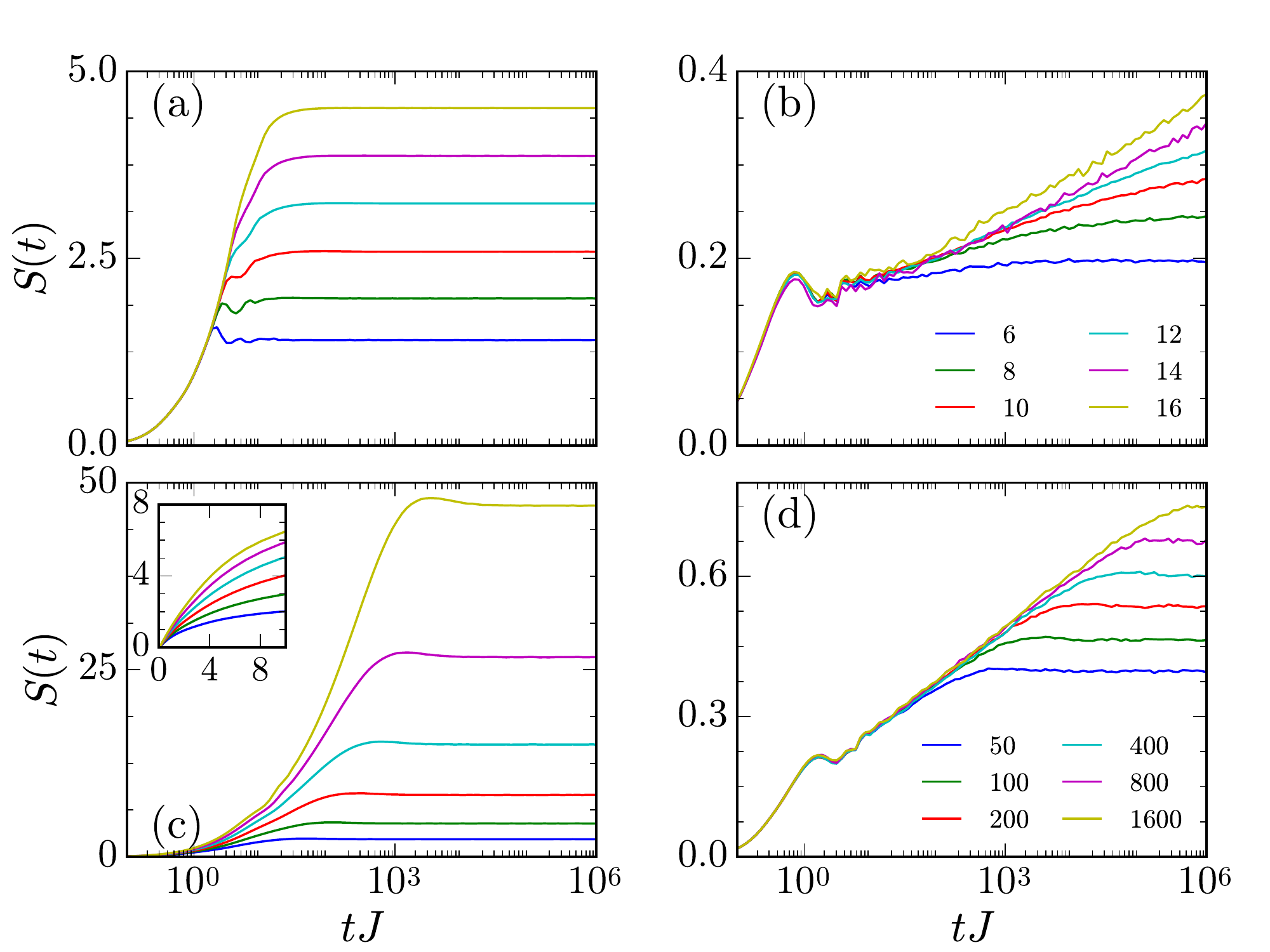}
\end{center}
\caption{(color online).
Comparison of entanglement dynamics between a long-range noninteracting model ($J \ne 0, ~U=0$) and a short-range interacting model ($\alpha,\beta \to \infty, ~J,U \ne 0$). Here (a) interacting delocalized phase ($\eta \approx J$) and (b) many-body localized phase ($\eta \approx 20J$) of the short-range interacting model, and (c) $\alpha=1,~\eta \approx 20J$ and (d) $\alpha=2,~\eta \approx 20J$ of the long-range noninteracting model.
The chain lengths are $6,8,\cdots,16$ in (a,b) while those are $50,100,200,\cdots,1600$ in (c,d).
In the inset of (c) we zoom in to show the length dependence at short times of the long-range model.
}
\label{XXZComparison}
\end{figure}
In this appendix, we consider $\alpha,\beta \to \infty$, which corresponds to short-range hopping and interactions. This short-range Hamiltonian in Eqn.~\ref{HamiltonianLongRange} for $\alpha,\beta \to \infty$ can be mapped by the Jordan-Wigner transformation to a 1D random-field XXZ spin chain which has been widely studied recently in the context of MBL~\cite{Pal2010,Bardarson2012,Roy2015,Singh2015}, and shows an MBL transition with increasing disorder strength. 

We show $S(t)$ in the interacting delocalized and MBL phases of the short-range model in the upper panel of Fig.~\ref{XXZComparison}.
One can immediately see striking differences in $S(t)$ between the two cases.
The delocalized phase shows a rapid (ballistic) growth of entanglement until it saturates at a rather large value satisfying a volume law.
In sharp contrast, $S(t)$ in the MBL phase grows slowly (logarithmically) over a very long time scale after a rapid initial growth.
This latter growth in this short-range model is entirely due to interactions, as it is absent in the AL phase~\cite{Bardarson2012,Serbyn2013,Singh2015}.
The logarithmic growth results in an extensive behavior (volume law) of the asymptotic entanglement $S_\infty$ in the MBL phase~\cite{Vosk2013,Singh2015}.
We compare the above two different behaviors of $S(t)$ in the short-range interacting model to that in the LRNM ($J \ne 0, ~U=0$) at $\alpha=1,2$ (see lower panel of Fig.~\ref{XXZComparison}).
They show interesting similarities. 
A rapid growth of $S(t)$ followed by system-size dependent saturation is observed in the delocalized phase of the short-range interacting model and the LRNM at $\alpha=1$. 
There is however one qualitative difference between Fig.~\ref{XXZComparison}(a) and Fig.~\ref{XXZComparison}(c), which is not immediately clear from the figure.
In the delocalized phase of the short-range interacting model, the long-distance correlations develop causally (with some finite maximum group velocity $v_g^{\rm max}$) whose physical mechanism is the one given by Lieb and Robinson~\cite{Lieb1972}.
As a result the entanglement is independent of system size at short times.
In the LRNM this picture is no longer valid for $\alpha<2$ and correlations can develop instantly at arbitrary long distances by long-range hopping resulting in infinite $v_g^{\rm max}$~\cite{Eisert2013, Hauke2013, Cevolani2015, Buyskikh2016, Cevolani2016}.
This is manifested in the appearance of a system-size dependence in the entanglement right from the beginning (see e.g., inset of Fig.~\ref{XXZComparison}(c)).
The behavior of $S(t)$  in the LRNM at $\alpha=2$ in Fig.~\ref{XXZComparison}(d) looks remarkably similar to the logarithmic growth seen in the MBL phase in Fig.~\ref{XXZComparison}(b).
However, there are some important differences between these two cases.
The first one is the scaling of asymptotic entanglement entropy with system size.
While the asymptotic entanglement is extensive in the MBL phase~\cite{Vosk2013,Singh2015}, the LRNM at $\alpha=2$ shows a non-extensive behavior, $S_{\infty} \sim \ln L$ (see Fig.~\ref{SystemSizeDependence}(b)).
A related feature is that the logarithmic growth due to interactions continues for a much longer duration than that due to long-range hopping (compare Fig.~\ref{XXZComparison}~(b) to \ref{XXZComparison}~(d)).
Note that the longest chain size in the interacting case ($L=16$) is much smaller than the smallest chain length in the noninteracting case ($L=50$) considered here.

The underlying physical mechanisms for the two logarithmic growths are most likely different because of the following observation.
In the MBL phase of a short-range interacting model, only entanglement and closely related quantities such as Renyi entropies show logarithmic growth.
In particular, particle number fluctuations $\mathcal{F}(t)$ do not show any signature of this growth~\cite{Bardarson2012,Singh2015}.
On the other hand, many physical quantities including $\mathcal{F}(t)$ show the logarithmic growth in the LRNM at $\alpha=2$.
This observation  suggests that a simpler physical mechanism involving actual movement of particles is causing the slow growth in the LRNM as opposed to the subtle dephasing mechanism between distant localized regions for the MBL case whose effect is seen only in suitable entropies~\cite{Serbyn2013, Huse2013}.
\section{Level-spacing and eigenstate thermalization in the long-range interacting model ($J \ne 0, ~U \ne 0$)}
\label{App2}
\begin{figure}
\begin{center}
\includegraphics[width=0.99\linewidth]{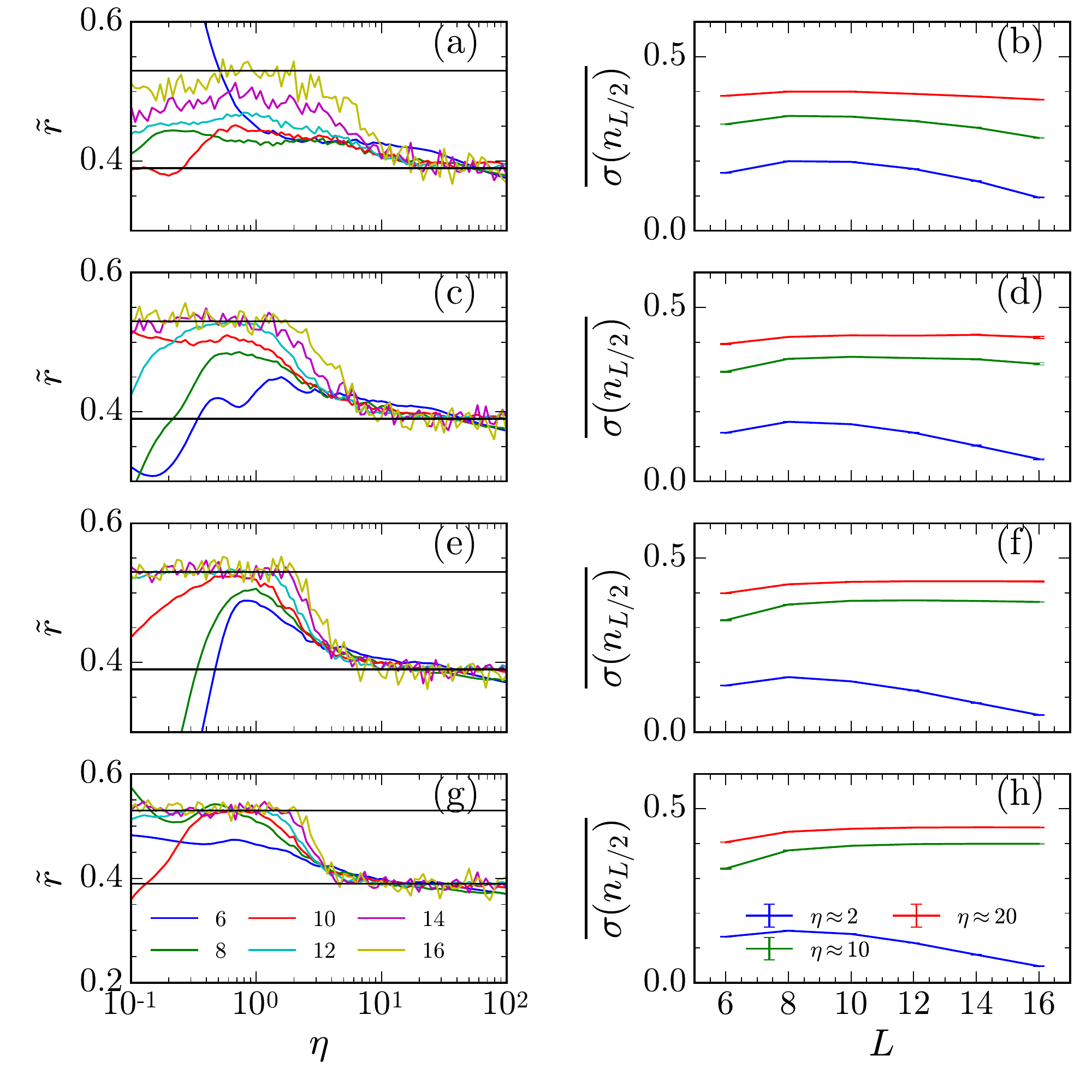}
\end{center}
\caption{(color online).
Left column: Localization-delocalization as evidenced by the level-spacing ratio $\tilde{r}$ as a function of disorder strength $\eta$. 
Different rows correspond to growing $\alpha=1,1.5,2,3$ 
from top to bottom, with different curves in each panel representing different system sizes, $L$. 
A localization-delocalization transition is clearly evident for all $\alpha > 1$.
The two solid lines in each panel show the GOE ($\approx 0.536$) and the Poisson ($\approx 0.386$) values for $\tilde{r}$.
Right column: thermalization and its breakdown. 
Shown is the disorder-averaged standard deviation of eigenstate expectation values of a local observable as a function of system-size $L$ for different values of disorder strength $\eta$ (different colors) and $\alpha=1,1.5,2,3$
from top to bottom. The local observable used is the number operator in the middle of the chain $n_{L/2}$ and $10\%$ eigenstates near the middle of the spectrum are used to calculate the standard deviation. For $\alpha>1$, a system-size independent standard deviation at large $L$ indicates breakdown of eigenstate thermalization.
}
\label{ThermalizationInteracting}
\end{figure}

In the first column of Fig.~\ref{ThermalizationInteracting} we show the level-spacing ratio $\tilde{r}$ in the middle of the spectrum of the long-range interacting model at different values of $\alpha$ ($\beta=1$).
Similar to other eigenstate properties, $\tilde{r}$ also appears to be relatively insensitive to the value of $\beta$ (except when $\alpha=1$), hence we here present results only for $\beta=1$.
We find a clear signature of delocalization-localization transition with increasing disorder for larger values of $\alpha$ with $\tilde{r}$ changing from the GOE value to its Poisson value.
Although a transition is not very clear for $\alpha=1, \beta=1$ we still see a Poisson level spacing for larger disorder.
The $\tilde{r}$ data appears to converge to the usual delocalization-localization behavior with increasing system size $L$ for higher values of $\alpha$ but it has not converged by $L=16$ for $\alpha=1, \beta=1$.
We explicitly test for eigenstate thermalization in the long-range interacting model by measuring fluctuations in the expectation value of a local operator over neighboring eigenstates~\cite{Beugeling2013}.
We choose the number operator $n_{L/2}$ on a site in the middle of the chain and find its expectation value.
We calculate the variance 
$$\sigma^2(n_{L/2}) = {\sum_e}' \left( \langle e|n_{L/2}| e \rangle \right)^2 -  \left( {\sum_e}' \langle e|n_{L/2}| e \rangle \right)^2$$
where the sum is over $10\%$ of eigenstates ($|e\ra$) near the middle of the spectrum.
In the second column of Fig.~\ref{ThermalizationInteracting} we show disorder averaged standard deviation $\overline{ \sigma(n_{L/2}) }$ as a function of system size $L$ for different values of disorder strength $\eta$ and the exponent $\alpha$.
When a system thermalizes, this is expected to decrease with increasing $L$ as can be seen for small disorder (delocalized phase) at all values of $\alpha$.
At a strong disorder it is roughly independent of $L$ which is a signature of the violation of eigenstate thermalization.
The behavior at $\alpha=1$ is more complicated as it appears that the fluctuation might decrease upon increasing $L$ further even at a large disorder.
This is consistent with the shift in MBL transition point to larger values of disorder with increasing system-size at $\alpha=1$ (Fig.~\ref{timeEvolutionInteracting}~d).
Higher values of $\alpha (>1)$ show thermalization (delocalized phase) and its breakdown (MBL phase) for the accessible system sizes. 

\section{Particle number fluctuations vs dynamical entanglement in the long-range interacting model ($J \ne 0, ~U \ne 0$)}
\label{App3}
\begin{figure}
\begin{center}
\includegraphics[width=0.99\linewidth]{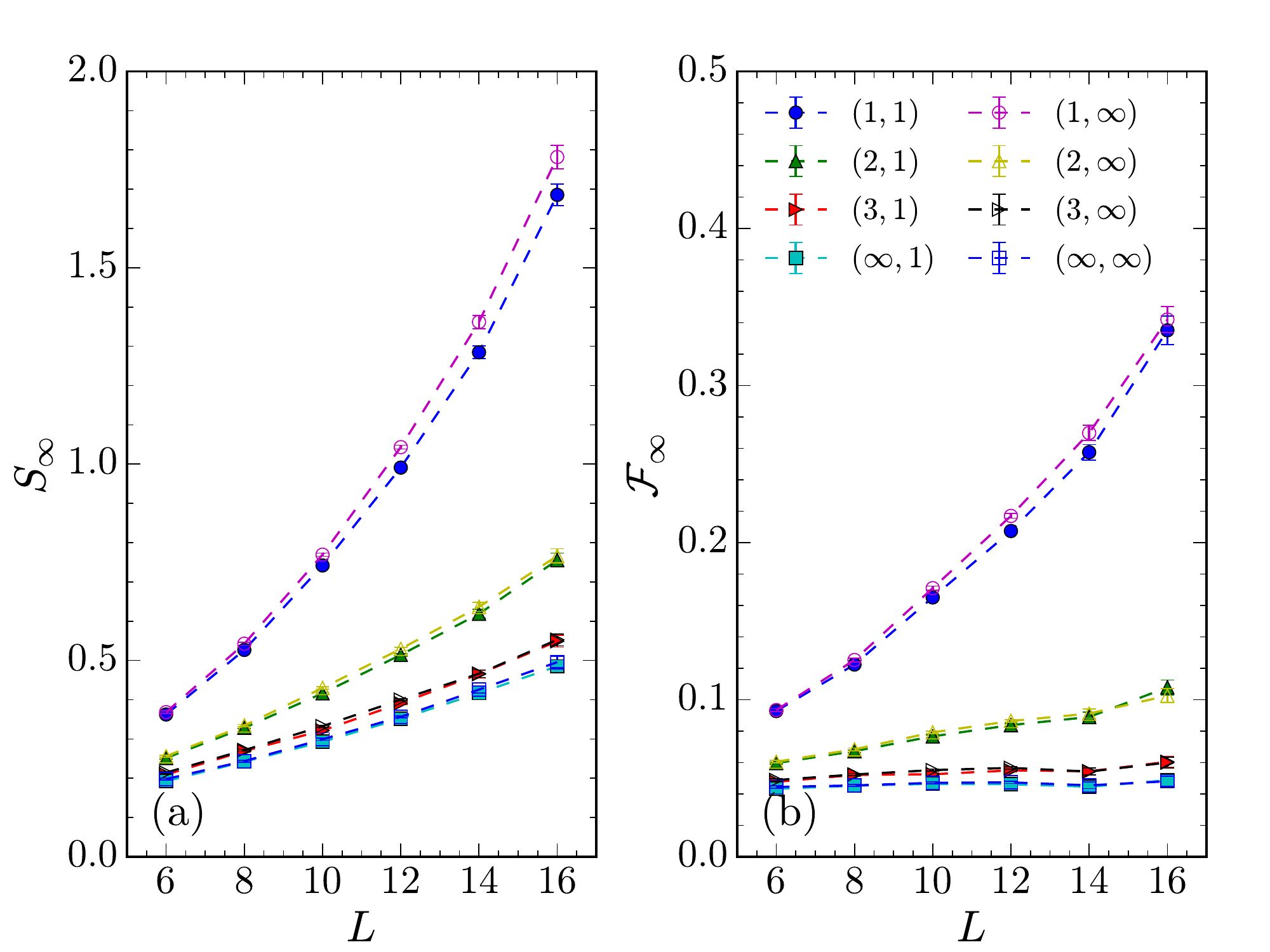}
\end{center}
\caption{(color online).
System size dependence of asymptotic dynamical entanglement ($S_\infty$) and particle number fluctuations ($\mathcal{F}_\infty$) in the long-range interacting model for $\eta \approx 20J$ at different values of exponents ($\alpha,\beta$).
}
\label{SystemSizeDependenceInteracting}
\end{figure}

The particle number fluctuations $\mathcal{F}(t)$ show interesting comparisons to entanglement in the long-range interacting model. We know from earlier studies that $\mathcal{F}(t)$ does not show slow-growth in the MBL phase for short-range interactions~\cite{Bardarson2012,Singh2015}, and we have mentioned in Sec. \ref{SectionXXZComparison} that  $\mathcal{F}(t)$ behaves qualitatively similar to $S(t)$ for the LRNM. Here we observe that $\mathcal{F}(t)$ does not follow the features of $S(t)$ in a long-range interacting model with short-range hopping. We show $L$-dependence of both $S_{\infty}$ and $\mathcal{F}(L,t\to \infty) \equiv \mathcal{F}_{\infty}$ for a strong disorder at different values of $\alpha$ and $\beta$ in Fig.~\ref{SystemSizeDependenceInteracting}.
We find that entanglement shows a clear $L$-dependence for all values of $\alpha$ and $\beta$.
On the other hand the $\mathcal{F}_{\infty}$  becomes independent of system size at large $\alpha$ irrespective of $\beta$.
The $L$-dependence of $S_{\infty}$ and $\mathcal{F}_{\infty}$ at small $\alpha$ appears to be faster than linear which is clearly a finite-size effect. The asymptotic entanglement is bounded by a linear dependence on $L$, $S_{\mathrm{max}} \le L \ln 2 / 2$, and a growth which is faster than linear would exceed the maximum allowed value for sufficiently long chains.

\bibliography{bibliography}
\end{document}